\documentclass[aps,prd,twocolumn,superscriptaddress,nofootinbib,showpacs,preprintnumbers]{revtex4}
\usepackage[english]{babel}

\begin{document}

\title{Will Black Holes eventually engulf the Universe?}
\author{Prado Mart\'{\i}n-Moruno}
\email{pra@imaff.cfmac.csic.es}
\affiliation{Colina de los Chopos, Instituto de Matem\'aticas y F\'\i sica Fundamental, \\ Consejo Superior de Investigaciones Cient\'\i ficas,
Serrano 121, 28006 Madrid, Spain}

\author{Jos\'e A. Jim\'enez Madrid}
\email{madrid@iaa.es}
\affiliation{Colina de los Chopos, Instituto de Matem\'aticas y F\'\i sica Fundamental, \\
Consejo Superior de Investigaciones Cient\'\i ficas,
Serrano 121, 28006 Madrid, Spain}
\affiliation{Instituto de Astrof\'\i sica de Andaluc\'\i a, Consejo Superior de
Investigaciones Cient\'\i ficas, Camino Bajo de Hu\'etor 50, 18008 Granada, Spain}
\author{Pedro F. Gonz\'alez-D\'\i az}
\email{p.gonzalezdiaz@imaff.cfmac.csic.es}
\affiliation{Colina de los Chopos, Instituto de Matem\'aticas y F\'\i sica Fundamental, \\
Consejo Superior de Investigaciones Cient\'\i ficas,
Serrano 121, 28006 Madrid, Spain}
\date{\today}

\begin{abstract}
The Babichev-Dokuchaev-Eroshenko model for the accretion of dark energy onto
black holes has been extended to deal with black holes with
non-static metrics. The possibility that for an asymptotic observer a
black hole with large mass will rapidly increase and eventually engulf
the Universe at a finite time in the future has been studied by using
reasonable values for astronomical parameters. It is concluded that such
a phenomenon is forbidden for all black holes in quintessential cosmological models.
\end{abstract}

\preprint{IMAFF-RCA-06-04}
\maketitle

The discovery of the current acceleration of the Universe has driven a
plethora of theoretical models  to account for the dark stuff
which has been invoked
to justify observations of distant supernovae Ia, cosmic microwave
background anisotropy, microlensing and the statistics of quasars and
clusters \cite{uno, dos, tres}, all aiming at determining the
equation-of-state parameter $w$ of the Universe, $p=w\rho$. It
seems still possible and even quite feasible that the value
of $w$ be less than $-1$, even
though this would induce serious problems for the consistency and
stability of  cosmic dark
energy \cite{cuatro,cinco,phantom3,phantom2,SSD,Li}.
However, even if $w>-1$, for which most of the above problems are no
longer present, the most popular dark energy model, i. e. the
quintessence scenario, may pose a serious potential difficulty. That
difficulty appears when we consider the accretion of a quintessence
scalar field with constant parameter $w>-1$ onto black
holes \cite{Jacobson:1999vr,Bean:2002kx,Frolov:2002va} . In fact, in
the simplest case first discovered by Babichev, Dokuchaev and
Eroshenko \cite{Babichev:2004yx,Babichev:2005py}, the rate of
Schwarzschild black hole mass is given by
\begin{equation}\label{uno}
\dot{M}\equiv\frac{{\rm d}M}{{\rm d}t}=4\pi A M^2[p(\rho_{\infty})+\rho_{\infty}],
\end{equation}
where $p(\rho_{\infty})$ and $\rho_{\infty}$ respectively are the pressure
and energy density of the Universe at the asymptotic
limit $r\rightarrow\infty$ and, in this case, $A=4$ \cite{Babichev:2004yx}.
One can then see from the onset that for $p+\rho>0$ and constant $w$
the increase of black hole mass may be so quick as to yield a black hole
mass corresponding to a size that would eventually exceed the size of
the Universe itself, at a finite time in the future $t=t_{bs}$, with
\begin{equation}
t_{bs}=t_0+\frac{1}{[4\pi A \rho_0M_0-(6\pi\rho_0)^{1/2}](1+w)}.
\end{equation}
It can be checked that such a rather weird behaviour would also occur in
the case of Kerr black holes \cite{JimenezMadrid:2005rk}.

Nevertheless, the whole extreme black hole swelling phenomenon could simply
be thought to be an artifact resulting from the use of static metrics
which do not induce any nonzero $\Theta^{r}_0$ component of the black
hole momentum-energy tensor, and hence implying no internal energy
flow. In this case, the accretion procedure used to derive the rate is
in principle only valid for small accretion rates, not for large rates
and even less for the extreme rates leading to a blow-up of the black hole
size. Our first task therefore will be checking under what conditions, if
any, the above catastrophic phenomenon may take place. The simplest
non-static metric that would still contain a time-dependence enough to
induce internal nonzero energy-flow component $\Theta^{r}_0$ to overcome
the above approximation on the rate is one in which the black-hole mass
is allowed to depend generically on time, that is for a Schwarzschild
metric, (we use natural units so that $G=c=1$):
\begin{eqnarray}\label{dos}
{\rm d}s^2&=&\left(1-\frac{2M(t)}{r}\right){\rm d}t^2-\left(1-\frac{2M(t)}{r}\right)^{-1}{\rm d}r^2{}
\nonumber\\
&&{}-r^2\left({\rm d}\theta^2+\sin^2\theta {\rm d}\phi^2\right).
\end{eqnarray}
By generalizing the mechanism advanced
in Refs. \cite{Babichev:2004yx,Babichev:2005py} to this non-static
case, after integrating the conservation laws for momentum-energy and
its projection on four-velocity, one can obtain first
\begin{eqnarray}\label{tres}
&(p+\rho)\left(1-\frac{2M}{r}+u^2\right)^{1/2}\times \nonumber\\ &ur^2M^{-2}e^{\int_{\infty}^{r}f(r,t){\rm d}r}=C_1(t),
\end{eqnarray}
where $C_1(t)$ is a time dependent function which has the physical
dimensions of an energy density, and
\begin{equation}\label{tresb}
f(r,t)=\frac{\partial_0T^0_0}{T^r_0}-\frac{4\pi r}{1-2M/r}\left(T^0_0-T^r_r\right),
\end{equation}
and secondly
\begin{equation}\label{cuatro}
ur^2e^{\int_{\rho_{\infty}}^{\rho}\frac{{\rm d}\rho}{p(\rho)+\rho}}e^{\int_{\infty}^{r}g(r,t){\rm d}r}=-A(t),
\end{equation}
where
\begin{eqnarray}\label{cuatrob}
g(r,t)&=&\frac{(1-2M/r+u^2)^{1/2}}{(1-2M/r)u}\frac{\partial_0\rho}{p+\rho}\nonumber\\
& &+\frac{\dot{M}(1-2M/r+2u^2)}{ur(1-2M/r)^2(1-2M/r+u^2)^{1/2}}\nonumber \\
& &+\frac{\partial_0u}{(1-2M/r)(1-2M/r+u^2)^{1/2}} \ \ \ ,
\end{eqnarray}
with $\partial_0\equiv \partial/\partial_t$ and $A(t)$ is a positive
time-dependent function (note that $u<0$). In obtaining
Eqs. (\ref{tres})-(\ref{cuatrob}) we have used for dark energy an
energy-momentum tensor $T^{\mu}_{\nu}$ corresponding to a perfect
fluid. In order to evaluate the function $A(t)$ we can use the property
that $A(t)={\rm lim}_{r\rightarrow\infty} ur^2$, so that $A(t)$ is not an
explicit function of $t$, because $r$ does not depend on $t$ for the
simplest non-static metric (\ref{dos}), and $u$ depends only on $t$
through $M$. Then, as the physical dimensions of $A(t)$ must
be ${\rm(meters)^2}$, we must have $A(t)=M^2A'$.
From Eqs. (\ref{tres}) and (\ref{cuatro}), it then follows:
\begin{eqnarray}\label{cinco}
&(p+\rho)\left(1-\frac{2M}{r}+u^2\right)^{1/2}\times \nonumber\\
&e^{-\int_{\rho_{\infty}}^{\rho}\frac{{\rm d}\rho}{p(\rho)+\rho}}e^{-\int_{\infty}^{r}{\rm d}r[g(r,t)-f(r,t)]}=B(t),
\end{eqnarray}
where $B(t)=-C_1(t)/A'=p(\rho_{\infty})+\rho_{\infty}$.

\vspace{24pt}
Relative to the energy flow induced in the quintessential fluid, the black
hole mass rate can be expressed as $\dot{M}=-\int{T^r_0 {\rm d}S}$, in
which $\mathrm{d} S=r^2\sin\theta\mathrm{d}\theta\mathrm{d}\phi$,
$0\leq\theta\leq\pi$ and $0\leq\phi\leq 2\pi$. Then
the above equations lead finally to
\begin{equation}\label{seis}
\dot{M}=4\pi A'M^2(p+\rho)e^{-\int_{\infty}^{r} f(r,t){\rm d}r}.
\end{equation}
One can see that for the relevant physical case of an asymptotic
observer; i.e. $r\rightarrow\infty$, one recovers Eq. (\ref{uno}) also for
the non-static black hole metric given by Eq. (\ref{dos}), such as it
occurs in the case of wormholes \cite{Pedro}. We note furthermore that
this asymptotic result is also valid for the Kerr-Newman black hole as it
has been shown \cite{JimenezMadrid:2005rk} that the effects due to the
presence of angular momentum and electric charge on the accretion of
dark energy all vanish as $r\rightarrow\infty$. It follows that, at least
asymptotically, the big swelling of black holes leading to a swallowing
of the Universe is not an artifact coming from metric staticity, although
for finite values of the radial coordinate, we see that the rate
equation is changed in such a way as to not guarantee the occurrence
of an extreme black hole swelling.

However, once we have shown the asymptotic consistency and
accuracy of Eq. (\ref{uno}), we must check how current observational
data fit the extent of this phenomenon, or even ultimately prevent it
from occurring at all.
Thus, integrating expression (\ref{dos}), we have
\begin{equation}\label{siete}
M=\frac{M_0}{1-F(t)},
\end{equation}
in which
\begin{equation}
F(t)=\frac{4\pi A'\rho_0 M_0(1+w)(t-t_0)}{1+(6\pi\rho_0)^{1/2}(1+w)(t-t_0)}.
\end{equation}
One can note then that $F(t)>0$ when $t>t_0$
and $\dot{F}(t)\equiv {\rm d}F(t)/{\rm d}t>0$; therefore $F(t)$ is an
increasing function and the limit of $F(t)$ when $t$ goes to infinity
is finite and equal to $(8\pi\rho_0/3)^{1/2}A'M_0$. Then $F(t)$ has a
asymptote. When we consider a universe model which contains only
dark energy (which is assumed to be a sufficiently good approximation for
the future times where the big swelling of the black holes could be thought to
occur), $\rho_0$, the energy density at the coincidence time, is given by
\begin{equation}
\rho_0=\rho_{n}(a_{n}/a_0)^{3(1+w)}=\rho_{n}(1+z)^{3(1+w)},
\end{equation}
in which the subindex $n$ indicates present value.
On the other hand, since the spatial curvature of the Universe $k$ is
thought to be nearly zero, the current density will be equal to the
current critical density, that is
\begin{equation}
\rho_{n}=\frac{3H^2_{n}}{8\pi}.
\end{equation}

Now, from the WMAP data, we can obtain
that $(8\pi\rho_n/3)^{1/2}=H_n\sim 10^{-26}{\rm(meters)^{-1}}$. One can
express the initial black hole mass
as $M_0=X_0M_{\odot}\sim X_0 10^3 {\rm(meters)}$, so that the asymptotic value
is $\sim X_0 10^{-23}<<1$, generally. In obtaining this estimate we have taken
into account that the value of constant $A'$ should be the same as that
for $A$ in Eq. (\ref{uno}) (see \cite{Babichev:2004yx}).
Thus, the main conclusion of the present paper is that for current
observational data and for solar-mass or even supermassive black holes
with masses even larger than $10^{10}M\odot$ the accretion of dark energy
with $w>-1$ onto black holes leads to a very smooth increase of the hole
size and the phenomenon of an engulfing of the Universe by the black
hole is largely prevented. In fact, it could never occur. We note that it
is only for extremely massive hypothetical black holes with masses on the
range of $M\sim 10^{23}M\odot$ or larger, that such a phenomenon could
eventually take place. It could then be possible that if $w$ kept constant
and accretion of ordinary matter would continue enlarging black holes in
the future, then these black holes might finally increase larger than
the Universe itself due to dark energy accretion.

Clearly, present observational indications seem to imply that $w$ is not
constant and takes on values less than $-1$, so that in principle, the
occurrence of the considered catastrophic phenomenon becomes actually quite
unlikely at any time in the far future. However, if $\dot{w}>0$ and dark
energy would therefore cross the dividing barrier at $w=-1$ in the
future, then the following speculative reasoning might be in order.

If the Universe will expand forever induced by dark energy with $w>-1$, then
it would commonly be believed than in about hundred thousand billion years
the last stars will die out, some of them leaving a black hole behind. These
black holes would evaporate by the Hawking process while the remaining
matter very slowly decayed. Thus both the cooled stars and the dilute gas,
and later the black holes formed at the end of the star lifetime and those
supermassive ones that stood at the center of galaxies, will disappear from
the Universe in its remote future, leaving rare electrons and positrons
spread over huge distances from each other. An unsolved question is
however, how big can a black hole grow? By the holographic bound we know
that black holes are the most entropic objects and by the laws of black hole
mechanics that coalescence of black holes implies a neat increase of
entropy. It appears then quite plausible, in principle, that superimposed to the above
process leading to the thermal death of the Universe, before
evaporating, black hole would continue swallowing ordinary and dark matter
or finally eating each other so that the total balance of entropy increase
be optimized to a maximum and actually the evaporation process will
effectively start with immensely huge black holes. However, if we assume that the Universe is expanding in size at the speed of light, then its radius would be 13.7 billion light years, and its diameter would be 27.4 billion light years. Converting this to meters and cubing gives a volume of $2\times 10^{30}$kg. This gives a total of $10^{23}$ stars. Thus, roughly speaking, for a black hole to undergo the big swelling leading to engulf the entire Universe we would need to start with an initial black hole mass which equals the mass of the entire Universe. This is nevertheless impossible because of the holographic bound for entropy, $S<S_{bh}$, where $S_{bh}$ would be the entropy of a black hole with the same energy as the whole Universe, and the associated feature that receding black holes at large distancy can never coalerse. It follows that in quintessence models black holes can never engulf the Universe.

\acknowledgements
The authors thank Salvador Robles and Alessandro Mantelli for useful discussions, technical help and encouragement. This work was
supported by MEC under Research Project No.FIS2005-01181.


\begin{thebibliography}{99}
\bibitem{uno}
  D.~J.~Mortlock and R.~L.~Webster,
  {\it Mon.\ Not.\ Roy.\ Astron.\ Soc.}\  {\bf 319} 872 (2000)
  [arXiv:astro-ph/0008081].
\bibitem{dos}
  A.~G.~Riess {\it et al.}  [Supernova Search Team Collaboration],
  {\it Astron.\ J.\ }  {\bf 116} 1009 (1998)
  [arXiv:astro-ph/9805201];
  S.~Perlmutter {\it et al.}  [Supernova Cosmology Project Collaboration],
  {\it Astrophys.\ J.\ } {\bf 517} 565 (1999)
  [arXiv:astro-ph/9812133];
   J.~L.~Tonry {\it et al.}  [Supernova Search Team Collaboration],
  {\it Astrophys.\ J.\ }  {\bf 594} 1 (2003)
  [arXiv:astro-ph/0305008].
 \bibitem{tres}
  D.~N.~Spergel {\it et al.}  [WMAP Collaboration],
  {\it Astrophys.\ J.\ Suppl.\ }  {\bf 148} 175 (2003)
  [arXiv:astro-ph/0302209];
   C.~L.~Bennett {\it et al.},
  {\it Astrophys.\ J.\ Suppl.\ }  {\bf 148} 1 (2003)
  [arXiv:astro-ph/0302207];
   M.~Tegmark {\it et al.}  [SDSS Collaboration],
  {\it Phys.\ Rev.\ } D {\bf 69} 103501 (2004)
  [arXiv:astro-ph/0310723].
\bibitem{cuatro}
  R.~R.~Caldwell,
  {\it Phys.\ Lett.\ } B {\bf 545} 23 (2002)
  [arXiv:astro-ph/9908168];
 A.~A.~Starobinsky,
  {\it Grav.\ Cosmol.\ }  {\bf 6} 157 (2000)
  [arXiv:astro-ph/9912054].
\bibitem{cinco}
  R.~R.~Caldwell, M.~Kamionkowski and N.~N.~Weinberg,
  {\it Phys.\ Rev.\ Lett.\ }  {\bf 91} 071301 (2003)
  [arXiv:astro-ph/0302506];
  P.~F.~Gonz\'alez-D\'{\i}az,
  {\it Phys.\ Lett.\  }B {\bf 586} 1 (2004).
\bibitem{phantom3}
B.~McInnes,
{\it JHEP }{\bf 0208} 029 (2002) [arXiv:hep-th/0112066];
S.~M.~Carroll, M.~Hoffman and M.~Trodden,
 {\it Phys.\ Rev.\ }D {\bf 68} 023509 (2003)
 [arXiv:astro-ph/0301273].
\bibitem{SSD} P.~Singh, M.~Sami and N.~Dadhich,
{\it Phys.\ Rev.\ }D {\bf 68} 023522 (2003)
[arXiv:hep-th/0305110];
J.~G.~Hao and X.~z.~Li,
{\it Phys.\ Rev.\ }D {\bf 70} 043529 (2004)
[arXiv:astro-ph/0309746];
\bibitem{Li} J.~g.~Hao and X.~z.~Li,
{\it Phys.\ Rev.\ }D {\bf 68} 043501 (2003)
[arXiv:hep-th/0305207];
{\it Phys.\ Rev.\ }D {\bf 68} 083514 (2003)
[arXiv:hep-th/0306033];
D.~j.~Liu and X.~z.~Li,
{\it Phys.\ Rev.\ }D {\bf 68} 067301 (2003)
[arXiv:hep-th/0307239].
\bibitem{phantom2}
P.~F.~Gonz\'{a}lez-D\'{i}az,
{\it Phys.\ Rev.\ }D {\bf 68} 021303 (2003)
[arXiv:astro-ph/0305559];
G.~W.~Gibbons,
arXiv:hep-th/0302199;
R.~Kallosh, J.~Kratochvil, A.~Linde,
E.~V.~Linder and M.~Shmakova,
{\it JCAP }{\bf 0310} 015 (2003)
[arXiv:astro-ph/0307185];
L.~P.~Chimento and R.~Lazkoz,
{\it Phys.\ Rev.\ Lett.\ } {\bf 91} 211301 (2003)
[arXiv:gr-qc/0307111]; M.~P.~Dabrowski, T.~Stachowiak and
M.~Szydlowski,
{\it Phys.\ Rev.\ }D {\bf 68} 103519 (2003)
[arXiv:hep-th/0307128];
V.~Faraoni,
{\it Phys.\ Rev.\ }D {\bf 68} 063508 (2003)
[arXiv:gr-qc/0307086];
H.~Stefancic,
{\it Phys.\ Lett.\ }B {\bf 586} 5 (2004)
[arXiv:astro-ph/0310904];
V.~B.~Johri,
{\it Phys.\ Rev.\ }D {\bf 70} 041303 (2004)
[arXiv:astro-ph/0311293];
J.~M.~Cline, S.~y.~Jeon and G.~D.~Moore,
{\it Phys.\ Rev.\ }D {\bf 70} 043543 (2004)
[arXiv:hep-ph/0311312];
 H.~Q.~Lu,
  {\it Int.\ J.\ Mod.\ Phys.\ } D {\bf 14}, 355 (2005)
  [arXiv:hep-th/0312082];
X.~H.~Meng and P.~Wang,
arXiv:hep-ph/0311070;
 S.~Nojiri and S.~D.~Odintsov,
{\it Phys.\ Lett.\ }B {\bf 571} 1 (2003)
[arXiv:hep-th/0306212];
{\it Phys.\ Lett.\ }B {\bf 562} 147 (2003)
[arXiv:hep-th/0303117];
{\it Phys.\ Lett.\ }B {\bf 565} 1 (2003)
[arXiv:hep-th/0304131];
{\it Phys.\ Lett.\ }B {\bf 599} 137 (2004)
[arXiv:astro-ph/0403622];
I.~Brevik, S.~Nojiri, S.~D.~Odintsov and L.~Vanzo,
{\it Phys.\ Rev.\ }D {\bf 70} 043520 (2004)
[arXiv:hep-th/0401073];
J.~S.~Alcaniz,
{\it Phys.\ Rev.\ }D {\bf 69} 083521 (2004)
[arXiv:astro-ph/0312424];
G.~Calcagni, 
{\it Phys.\ Rev.\ }D {\bf 69} 103508 (2004)
[arXiv:hep-ph/0402126];
J.~M.~Aguirregabiria, L.~P.~Chimento and R.~Lazkoz,
{\it Phys.\ Rev.\ }D {\bf 70} 023509 (2004)
[arXiv:astro-ph/0403157];
J.~G.~Hao and X.~Z.~Li,
{\it Phys.\ Lett.\ }B {\bf 606} 7 (2005)
[arXiv:astro-ph/0404154].





\bibitem{Jacobson:1999vr}
  T.~Jacobson,
  {\it Phys.\ Rev.\ Lett.\ }  {\bf 83}, 2699 (1999)
  [arXiv:astro-ph/9905303].


\bibitem{Bean:2002kx}
  R.~Bean and J.~Magueijo,
  {\it Phys.\ Rev.\ } D {\bf 66}, 063505 (2002)
  [arXiv:astro-ph/0204486].



\bibitem{Frolov:2002va}
  A.~V.~Frolov and L.~Kofman,
  {\it JCAP  }{\bf 0305}, 009 (2003)
  [arXiv:hep-th/0212327].
\bibitem{Babichev:2004yx}
  E.~Babichev, V.~Dokuchaev and Y.~Eroshenko,
  {\it Phys.\ Rev.\ Lett.\ }  {\bf 93} 021102 (2004)
  [arXiv:gr-qc/0402089].

\bibitem{Babichev:2005py}
  E.~Babichev, V.~Dokuchaev and Y.~Eroshenko,
  {\it Zh.\ Eksp.\ Teor.\ Fiz.\ }  {\bf 100} 597 (2005)
  [arXiv:astro-ph/0505618].
\bibitem{JimenezMadrid:2005rk}
  J.~A.~Jim\'enez Madrid and P.~F.~Gonz\'alez-D\'{\i}az,
  arXiv:astro-ph/0510051.
\bibitem{Pedro}
P.~F.~Gonz\'alez-D\'{\i}az,
``{\it Some notes on the Big Trip}'',
{\it Phys. lett. } B (in press, 2006);
P.~Mart\'{\i}n-Moruno, J.~A.~Jim\'enez Madrid and P.~F.~Gonz\'alez-D\'{\i}az,
in preparation.
\bibitem{termica}
I.D.Novikov, "{\it Whither Does the River of Time Flow?}", molodaya Gvardiya, Moscow, 1990 (in Russian). A partial English translation can be found in: I.D. Novikov, "{\it The River of Time}" (Cambridge University Press, Cambridge, UK, 1998);
P.C.W. Davies, "{\it The Last Three Minutes: Conjectures About the Ultimate Fate of the Universe}" (Basic Books, New York, USA, 1997)
\end{thebibliography}
\end{document}